\documentclass[runningheads]{llncs}
\usepackage{graphicx}
\usepackage{amsmath}
\usepackage{graphicx}
\usepackage{amssymb}  
\usepackage{xcolor}   
\usepackage{hyperref} 
\usepackage{csquotes}
\usepackage{subcaption}
\usepackage{multirow}
\usepackage[normalem]{ulem}
\usepackage[absolute,overlay]{textpos} 

\begin{document}
\title{
Analysis of the Scalability of a Deep-Learning Network for Steganography \enquote{Into the Wild}
}
\titlerunning{Analysis of the Scalability of a DL Network for Steganography}
%
\author{
Hugo Ruiz\inst{1}\orcidID{0000-0002-2806-2428} \and
Marc Chaumont\inst{1,2}\orcidID{0000-0002-4095-4410} \and \\
Mehdi Yedroudj\inst{1}\orcidID{0000-0001-6404-3876} \and 
Ahmed Oulad Amara\inst{1} \and \\
Frédéric Comby\inst{1}\orcidID{0000-0001-7157-4296} \and
Gérard Subsol\inst{1}\orcidID{0000-0002-7461-4932}
}
\authorrunning{H. Ruiz et al.}
%

\institute{
Research-Team ICAR, LIRMM, Univ. Montpellier, CNRS, France \\
\email{\{hugo.ruiz, marc.chaumont, mehdi.yedroudj \\ ahmed.oulad-amara, frederic.comby, gerard.subsol\}@lirmm.fr}
\and University of N\^{i}mes, France}
%
\maketitle              
\begin{abstract}

Since the emergence of deep learning and its adoption in steganalysis fields, most of the reference articles kept using small to medium size CNN, and learn them on relatively small databases.

Therefore, benchmarks and comparisons between different deep learning-based steganalysis algorithms, more precisely CNNs, are thus made on small to medium databases. This is performed without knowing: 
\begin{enumerate}
    \item if the ranking, with a criterion such as accuracy, is always the same when the database is larger,
    \item if the efficiency of CNNs will collapse or not if the training database is a multiple of magnitude larger,
    \item the minimum size required for a database or a CNN, in order to obtain a better result than a random guesser.
\end{enumerate}

In this paper, after a solid discussion related to the observed behaviour of CNNs as a function of their sizes and the database size, we confirm that the error's power-law also stands in steganalysis, and this in a border case, i.e. with a medium-size network, on a big, constrained and very diverse database.

\end{abstract}
%
%
%
\section{Introduction}
\label{sec:introduction}
\begin{textblock*}{122mm}(4.7cm,1cm)
{\footnotesize \textnormal{
\textit{ICPR'2021, International Conference on Pattern Recognition, MMForWILD'2021, Worshop on MultiMedia FORensics in the WILD, Lecture Notes in Computer Science, LNCS, Springer. January 10-15, 2021, Virtual Conference due to Covid (formerly Milan, Italy). Minor corrections added to the published version (29th of December 2020).\vspace{+0.7cm}\\}}}
\end{textblock*}

Steganography is the art of concealing information in a common medium so that the very existence of the secret message is hidden from any uninformed observer. Conversely, steganalysis is the art of detecting the presence of hidden data in such mediums \cite{Fridrich2009_Book}.

Since 2015, thanks to the use of Deep-Learning, steganalysis performances have significantly improved \cite{Chaumont2020}. Nevertheless, in many cases, those performances depend on the size of the learning set. It is indeed commonly shared that, to a certain extent, the larger the dataset, the better the results \cite{Yedroudj_pixelsoff}. Thus, increasing the size of the learning set generally improves performance while also allowing for more diverse examples during training. 
 
The objective of this article is to highlight the performance improvement of a Deep-Learning based steganalysis algorithm as the size of the learning set increases. In such a context, the behaviour of the network has never been studied before, and numerous questions related to model size and dataset size are still unsolved. 

In section \ref{sec:model_data_scaling}, we discuss those questions and the generic laws or models that have been proposed by the scientific community. Next, in Section \ref{sec:testbed}, we present the testbench used to assess the error power-law. We justify and discuss the various choices and parameters setting, required in order to run the experiments. We also present the Low Complexity (LC-Net) network \cite{Huang2019_LMC} which is a CNN that was considered as the state of the art algorithm for JPEG steganalysis in 2019 and 2020. In the experimental section \ref{sec:exp}, we briefly
present the Large Scale Steganalysis Database (LSSD) \cite{Ruiz2021_LSSD}, the experimental protocol and describe the conducted experiments. We then analyze the accuracy evolution with respect to the learning set size, and predict, thanks to the error power-law, the reachable efficiency for very big databases. Finally, we conclude and give some perspectives.

\section{Model scaling and Data scaling}
\label{sec:model_data_scaling}

Many theoretical and practical papers are trying to better understand the behaviour of neural networks when their dimension is increasing (the depth or the width) \cite{Belkin2019_DoubleDescent} \cite{Spigler_2019_under_over} \cite{Advani2020_GeneralizationError} \cite{Nakkiran2020_ICLR_Bigger_Hurt}, or when the number of examples is increasing \cite{Hestness2017} \cite{Sala2019_Power_Law} \cite{Rosenfeld2020_ICLR_Prediction-Across-Scales}. To this end, lots of experiments are done in order to observe the evolution of the test error as a function of the {\it model size}, or as a function of the {\it learning set size} and general laws are proposed for modelling the phenomenon. Those are essential researches because finding some generic laws could confirm that CNN users are applying the right methodologies.


Even if they have access to a large dataset, which is, in many domains, rarely possible, CNN users may have to restrict the learning to small to medium database, and small to medium-size models, during the preliminary experiments. Then, once satisfied, if possible, they can run a long time learning on a large dataset (i.e. greater than $10^6$ examples) and eventually with a large network (i.e. greater than $10^6$ parameters).

The questions arising by users are then: do the comparisons between various models, when evaluated on a small dataset, also stand when the dataset size increase. In other words, can we reasonably conclude on the best model when comparing the networks on a small dataset? What is the behaviour of a medium-size network when the dataset size increases? More generally, is there a collapse in performance when a model or the dataset scales up? Or, will the accuracy increase? Should we prefer bigger models? Is there a minimum required size for models or dataset?

\begin{figure}[h!]
	\centering
		\includegraphics[scale=0.45]{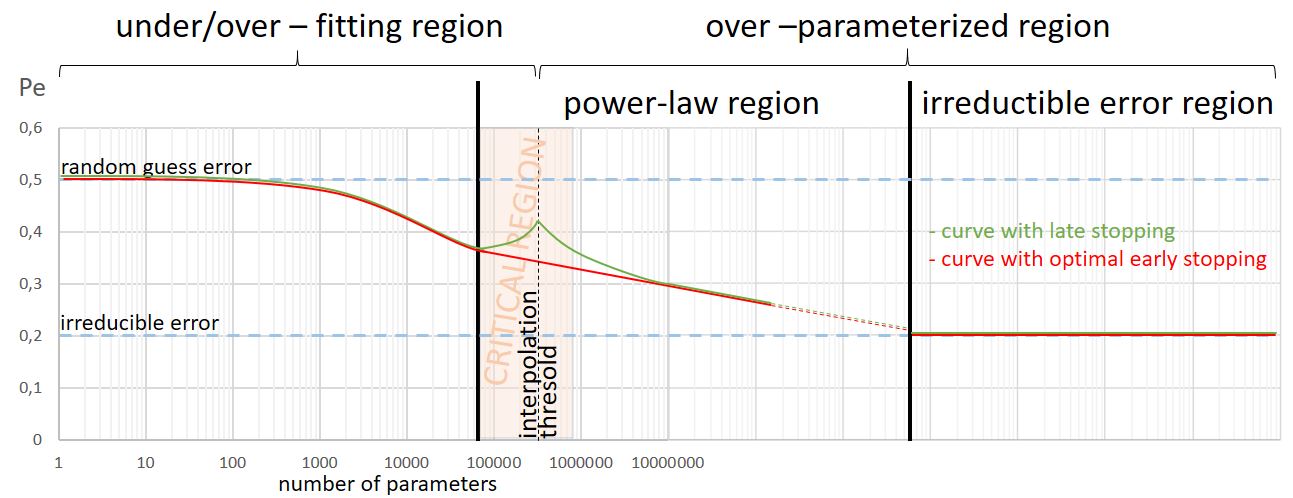}
		\caption{Schematic generic evolution of the test error depending on the model size.}
	\label{fig:scheme_model_law}	
\end{figure}
\begin{figure}[h!]
	\centering
		\includegraphics[scale=0.45]{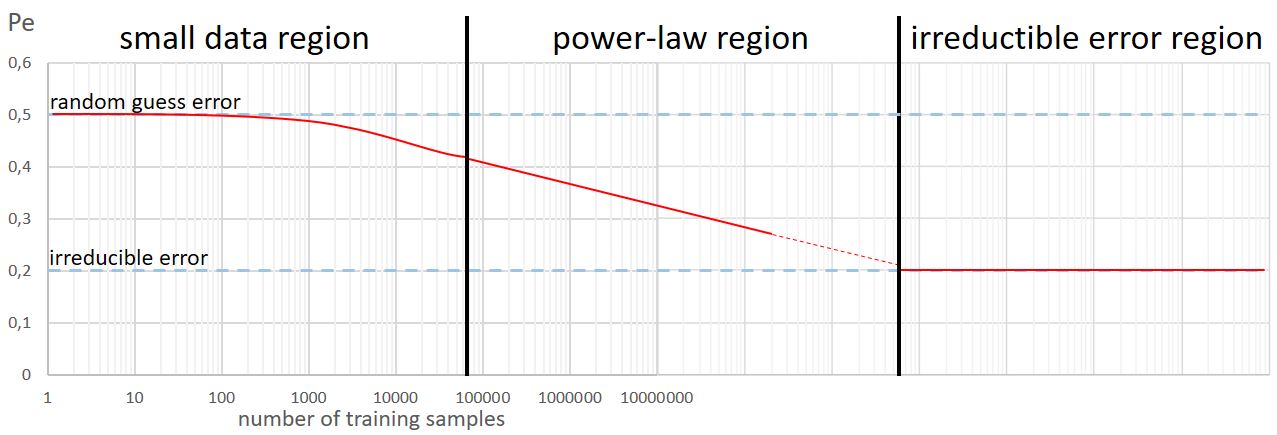}
		\caption{Schematic generic evolution of the test error depending on the dataset size.}
	\label{fig:scheme_data_law}
\end{figure}

In the studies related to the {\it model scaling}, researchers have observed three regions depending on the model size. There is the {\it underfitting}, the {\it overfitting}, and finally the {\it over-parameterized} region. The transition point to the over-parameterized region is named the {\it interpolation threshold}. Figure \ref{fig:scheme_model_law} shows those three regions (note that the abscissa scale is logarithmic).

When looking to the curve of the error as a function of the model size (the green curve; this without optimal early stopping), we can observe a {\it double descent} \cite{Nakkiran2020_ICLR_Bigger_Hurt}. In general, a practical conclusion is that the over-parameterized networks, i.e. with millions of parameters, can be used for any task and that it is beneficial using those more complex models. This idea has been, for example, used in practice in order to build gradually larger EfficientNet networks \cite{Tan2019_EfficientNet}. Note that this network has been strongly used by competitors \cite{Yousfi2020_Alaska2} \cite{Chubachi2020_Alaska2} during the Alaska\#2 competition \cite{Cogranne2020_Alaska2}.

In the studies related to the data scaling, researchers have observed that there are also three regions depending on the dataset size \cite{Hestness2017}. There is the {\it small data} region, the {\it power-law} region, and finally the {\it irreducible error} region. Figure \ref{fig:scheme_data_law} shows those three regions (note that the abscissa scale is logarithmic).

In the power-law region, the more data, the better results \cite{Sala2019_Power_Law},\cite {Yedroudj2018_DatabaseAugmentation}.

Recently, the authors of \cite{Rosenfeld2020_ICLR_Prediction-Across-Scales} have proposed a generic law that models the behaviour when scaling both the model size and the dataset size. Briefly, the test error noted $\epsilon$ is expressed as the sum of two exponentially decreasing term plus a constant. The first term is function of the dataset size, noted $n$, and the second one is function of the model size, noted $m$ \cite{Rosenfeld2020_ICLR_Prediction-Across-Scales}:
\begin{eqnarray}
\epsilon : \mathbb{R} \times \mathbb{R} & \rightarrow & [0, 1] \\
\epsilon(m,n) & \rightarrow &  
  \underbrace{a(m)n^{-\alpha(m)}}_{dataset \ power-law}
+ \underbrace{b(n)m^{-\beta(n)}}_{model \ power-law}
+c_{\infty} \nonumber
\label{eq::espilon}
\end{eqnarray} 
with $\alpha(m)$ and $\beta(n)$ controlling the rate of the error decrease, depending on $m$ and $n$ respectively, and $c_{\infty}$ the irreducible error, a real positive constant, independent of $m$ and $n$.

Then, the authors propose a simplification of the expression in:
\begin{equation}
\tilde{\epsilon}(m,n) = a n^{-\alpha} + b m^{-\beta}+c_{\infty}
\end{equation}
with $a$, $b$, $\alpha$, and $\beta$ real positive constants, and then use a complex envelope function in order to represent the transition from the {\it random-guess error} region to the {\it power-law} region \cite{Rosenfeld2020_ICLR_Prediction-Across-Scales}: $\hat{\epsilon}(m,n)=\epsilon_0\left\lVert \tilde{\epsilon}(m,n)/(\tilde{\epsilon}(m,n) -i\eta) \right\rVert$, with $\epsilon_0 =1/2$ for balanced binary classification, $i=\sqrt{-1}$, and $\eta\in \mathbb{R}$.

The interesting aspect with this function is that once the parameters $a$, $b$, $\alpha$, $\beta$, and $\eta$ are learnt using a regression on experimental points, obtained at various $m$ and $n$ values, with $m$ and $n$ not too high, one can answer to the questions mentioned above \footnote{See the paper \cite{Rosenfeld2020_ICLR_Prediction-Across-Scales}, and the discussions here: \url{https://openreview.net/forum?id=ryenvpEKDr}.}.

Now, let us go back to a more practical aspect. Suppose we are learning with an efficient network with enough parameters, i.e. on the right region relative to the interpolation threshold, possibly leaving the critical region (see Figure \ref{fig:scheme_model_law}), and use a data-set of medium size such that we are no more in the
small data region, avoiding us a random guess error (see Figure \ref{fig:scheme_data_law}). When studying the effect of increasing the data on the error, we should be in the power-law region and equation \ref{eq::espilon} can be simplified, as in \cite{Hestness2017}: 
\begin{equation}
\epsilon(n) = a' n^{-\alpha'} + c'_{\infty}
\label{eq::power-law-dataset-size}
\end{equation}

In the rest of our paper, we are observing, in the context of JPEG steganalysis, the behaviour of a medium network when the dataset size increases. Then, we confront these results to the power-law related to the data scaling (Equation \ref{eq::power-law-dataset-size}). 
Moreover, we are checking a “border case” because we are using a medium size model ($ 3.10^{5} $ parameters), and because we are using a very diverse database (the LSSD database \cite{Ruiz2021_LSSD} derived, from a part, from Alaska\#2 \cite{Cogranne2020_Alaska2}). This could result in a collapse in performance as the database increases, and failure to comply with the evolution law of the estimation error.


\section{A test bench to assess scalability for DL-based steganalysis}
\label{sec:testbed}

\subsection{Discussion on the test bench design}
	
{\hspace{0.5cm}{\bf Choice of the network:}	Our objective is to evaluate the accuracy (or equivalently the probability of error) as a function of the increase in the size of the dataset. But, as many researchers, we are limited by computational resources, so we need a low complexity network. We thus selected the Low Complexity network (LC-Net) \cite{Huang2019_LMC}, for its medium size (300 000 parameters), and its good performance as it is recognized as a state-of-the-art CNN for JPEG steganalysis at the date we ran the experiments (between September 2019 to August 2020). Note that we can consider that the LC-Net is probably close to the {\it interpolation threshold}\footnote{The ResNet18 with a width=10 is made of 300,000 parameters and is in the {\it interpolation threshold} region for experiments run on CIFAR-10 and CIFAR-100 in \cite{Nakkiran2020_ICLR_Bigger_Hurt}.} which implies that we must take caution to do an early stopping, close to the optimal, during the learning phase.
	
{\bf Choice of the payload:} Another critical thing is that the network should be sufficiently far from the random-guess region in order to observe a concrete improvement of the performance when scaling the database. So we have to choose a payload in order that the LC-Net accuracy is quite far from 50\%. This is quite challenging because there are no experiments results for LC-Net \cite{Huang2019_LMC} on "controlled" databases such as LSSD having a large diversity. More generally, there are not so many experiments reported before the summer of 2020 that used the unique controlled and diverse, Alaska\#1 \cite{Cogranne2019_Alaska} database.  
The objective, here, was to obtain accuracy between 60\% and  70\% for a small database (but not too small
\footnote{A too small database could bias the analysis since there is a region where the error increases when the dataset increase (see \cite{Nakkiran2020_ICLR_Bigger_Hurt}). For example, in \cite{Yedroudj2018_DatabaseAugmentation}, we report that the number of images needed, for the medium size Yedroudj-Net \cite{Yedroudj2018_Net}, to reach a region of good performance for spatial steganalysis (that is the performance of a Rich Model with an Ensemble Classifier), is about 10,000 images (5,000 covers and 5,000 stegos) for the learning phase, in the case where there is no cover-source mismatch, and the image size is 256$\times$256 pixels. 
}), in order to observe progression when the dataset is scaled and to let the possibility to future better networks to beat our results with sufficient margin. After a lot of experimental adjustments, either related to the building of the LSSD database \cite{Ruiz2021_LSSD} or related to the LC-Net \cite{Huang2019_LMC}, we found that 0.2 bpnzacs was a good payload for grey-level JPEG 256$\times$256 image with a quality factor of 75, ensuring to be quite far from the random-guess region for a small database of 20,000 images made of cover and stego images.

{\bf Choice related to the database:} We decided to work on grey-level JPEG images in order to put aside the color steganalysis, which is still recent and not enough theoretically understood  \cite{Abdulrahman2016_ColorSteganalysis}. Related to color steganalysis, the reader can consult the paper WISERNet \cite{Zeng2019_WISERNet} or the description of the winning proposition for Alaska\#1 \cite{Yousfi2019_BreakingAlaska}. The reader can also read the even more recent papers, in the top-3 of Alaska\#2 \cite{Yousfi2020_Alaska2}, \cite{Chubachi2020_Alaska2}, which are based on an ensemble of networks (for example EfficientNet \cite{Tan2019_EfficientNet}), which have preliminary learned on ImageNet. Related to steganography, the most recent proposition in order to take into account the three channels during the embedding can be found in \cite{Cogranne2020_DeLS-JPEG-Color} and was used in order to embed payload in Alaska\#2 images. 

We also decided to work only on the quality factor 75, and thus let apart the quantization diversity. Nevertheless, the conclusions obtained in the following could probably extend to a small interval around quality factor 75. Indeed, the authors of \cite{Yousfi2020_JPEG_QualityFactor} show that applying a steganalysis with a JPEG images database made of multiple quality factors could be done without efficiency reduction, using a small set of dedicated networks, each targeting a small interval of quality factor. Finally, it is maybe possible to use a unique network when there are various quality factors, as it has been done by a majority of competitors during Alaska\#2, thanks to the use of a pre-learned network on ImageNet. An extension of our work to a database with a variety of quality factors is postponed to future research.

Another reason to work with a quality factor of 75 is that we had in mind, for future work, the use of non controlled images such as ImageNet. ImageNet is made of JPEG compressed images whose development process is not controlled and whose quality factors are multiple. By re-compressing the images with a smaller quality factor, such as 75, the statistical traces of the first initial compression are removed. Such a re-compression would allow us to work on images with a roughly similar quality factor, and whose statistical properties would not be too poor. Additionally, the experimental methodology would be close to those exposed in the current paper and would facilitate comparisons.  

Finally, we built the LSSD database \cite{Ruiz2021_LSSD} in order to have a proper set for our experiments. For this database, we used controlled development using scripts inspired by Alaska\#2. The LSSD was obtained by merging 6 public RAW images databases including Alaska\#2. Without being as varied as images available on the Internet, we consider that diversity is nevertheless significant. There are 2 million covers in LSSD learning database, and we built "included subsets", from those 2 million covers in order to run our experiments.

With all those precautions, at the date where experiments started i.e. before running the experiment with an increasing number of examples, we assumed that we were at the border case, where the power-law on the data is valid.

		\subsection{Presentation of LC-Net}
In this paper, we use the Low Complexity network (LC-Net) \cite{Huang2019_LMC}, which is a convolutional neural network proposed in 2019 for steganalysis in the JPEG domain. Its design is inspired by ResNet \cite{He2016_ResNet}, the network that won the ImageNet competition in 2015. LC-Net performance is close to the state-of-the-art SRNet \cite{Boroumand2018_SRNet}, with the advantage of a significant lower complexity in terms of the number of parameters \cite{Huang2019_LMC} (twenty times fewer parameters than SRNet). This reduction in the number of parameters leads to less learning time as it converges faster toward an optimal solution. 

LC-Net is composed of three modules: pre-processing, convolution and classification (see Fig. \ref{fig:lc_net}).

The pre-processing module has a total of 4 convolutional layers, with the first layer kernels initialized using 30 SRM filters. These high-pass filters are commonly used in steganalysis \cite{Ye2017}, \cite{Yedroudj2018_Net}. They allow the network to reduce its learning time but also to converge when using a small learning set. For instance, using the BOSS database \cite{Bas2011-BOSS}, only 4,000 pairs of cover-stego images may be sufficient to perform learning \enquote{from scratch} and get good performance \cite{Yedroudj2018_DatabaseAugmentation}. The parameters of the first layer are not fixed; they are optimized during training. This first layer is followed by an activation function \enquote{TLU} (Truncated Linear Unit) \cite{Ye2017}, where the $ T$ threshold is set to $31$. For the remaining layers of the network, the \enquote{ReLU} (Rectified Linear Units) activation function is used. Batch normalization is used. No pooling is applied in this first module in order to preserve the stego signal.

The convolutional module is composed of 6 blocks, all with residual connections. These connections allow to avoid the vanishing gradient phenomenon during the back-propagation and thus to have deeper networks. The first two blocks have only two convolutional layers with direct residual connections. Blocks 3 to 6 include $3\times3$ convolutions with a stride equal to 2 to reduce the size of the feature maps. Indeed, it preserves the complexity of the computation time per layer when the number of filters is doubled \cite{He2016_ResNet}. Blocks 4 to 6, are Bottleneck residual blocks \cite{He2016_ResNet}. These blocks include 3 convolutional layers, a $1 \times 1$ convolution layer, a $ 3 \times 3$ convolution layer and another $ 1 \times 1$ convolution layer. The use of the Bottleneck block \cite{He2016_ResNet} allows the Low Complexity network having fewer parameters.

Finally, the classification module consists of a \enquote{Fully Connected} (FC) layer and a Softmax activation function. This function allows for obtaining the probabilities of the cover and stego classes.

\begin{figure}[h!]
	\centering

	\includegraphics[scale=0.64]{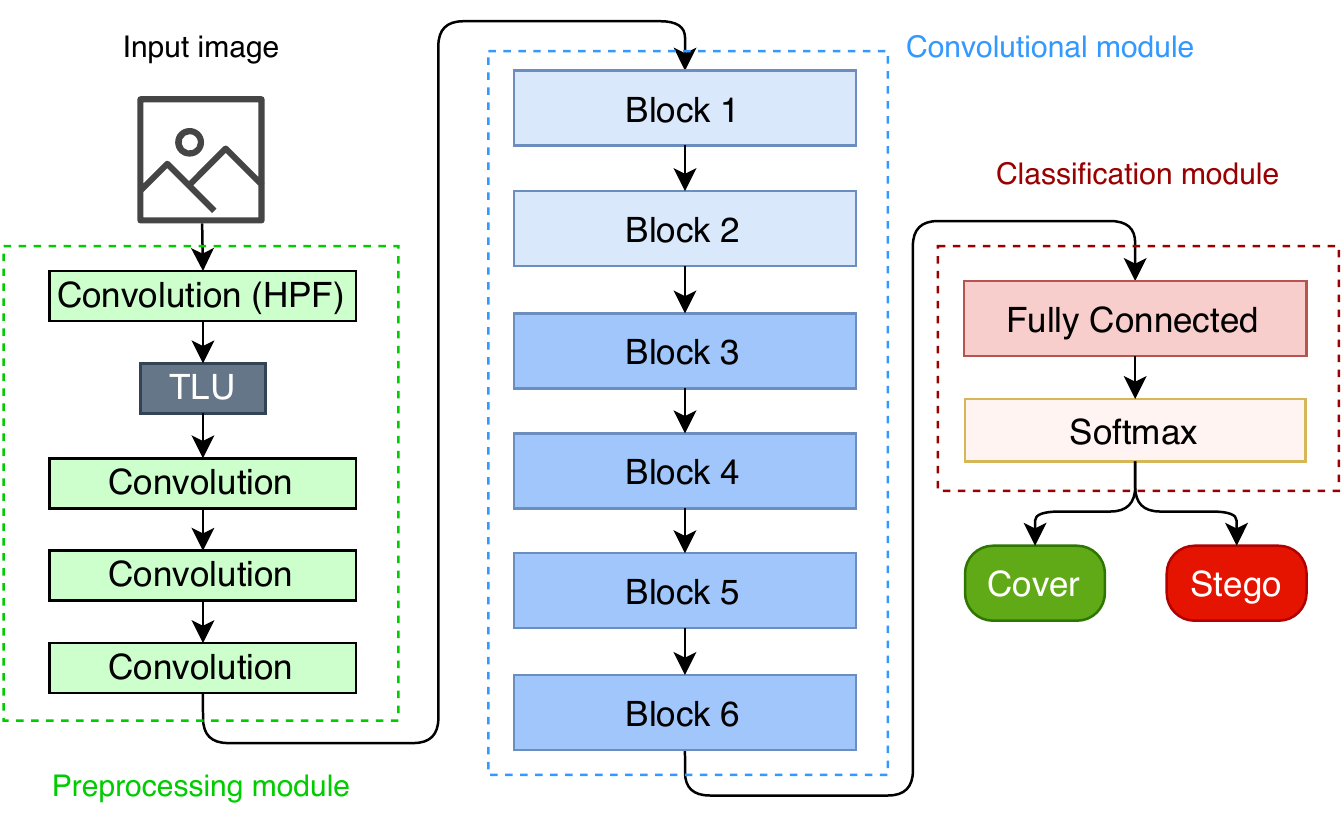}

	\caption{Architecture of LC-Net}
	\label{fig:lc_net}
\end{figure}

	\section{Experiments and results}
	\label{sec:exp}
		\subsection{Dataset and software platform}
As mentioned previously, the experiments were conducted on the LSSD database \cite{Ruiz2021_LSSD}\footnote{The LSSD database is available at: http://www.lirmm.fr/\~{}chaumont/LSSD.html.}. 
We are using greyscale JPEG images with a quality factor of 75. The size of those JPEG images is $256\times256$ pixels. They were obtained by developing RAW images (data issued from the camera sensors) from  ALASKA\#2, BOSS, Dresden, RAISE, Stego App, and Wesaturate public databases. The development scripts are inspired from the Alaska\#2 scripts. 

The cover database used for the learning phase is made of 2 million images. There is also a version with 1M, 500k, 100k, 50k, and 10k images. The 1M cover images database is a subset of the 2 million one, and so on, until the 10k cover images database. Each of those cover databases retains the same proportions of images from the different public databases. 

The cover database used for the test phase is made of 100k images and will always be the same whatever the experiments. This test database is obtained by developing RAW images which were not present in the learning cover database. The test cover database roughly keeps the distribution of the origins of the public databases. Thus, the steganalysis scenario is close to a clairvoyant scenario where the test set and learning set are statistically very close.

In our experiments, we have only used the 500k, 100k, 50k and 10k versions of the cover database due to excessively long learning time process for 1M and 2M images versions. 

The study was conducted on an IBM software container, with access to 144 supported POWER9 Altivec processors (MCPs) and two GV100GL graphic cards (Tesla V100 SXM2 16Gb).
		\subsection{Training, validation, and testing}
The embedding process has been done with the Matlab implementation of J-UNIWARD algorithm \cite{Holub2014_UNIWARD}, with a payload of 0.2 bits per non-zero AC coefficient (bpnzacs). It took almost three days (2 days and 20 hours) for the embedding on an Intel Xeon W­2145 (8 cores, 3.7­4.5 GHz Turbo, 11M cache).

Before feeding the neural network, JPEG images have to be decompressed in order to obtain spatial non rounded "real values" images. This essential step takes approximately 18 hours for all the images. Note that storage space requirement becomes important. Indeed, for a $256\times256$ grey-scale image, the file's size is around 500 kB when it is stored in MAT format in \textit{double} format. Thus, a database of 2M images requires a storage space of about 2 TB, and the learning cover bases, from 10k to 2M images, as well as the test cover database, in both JPEG and MAT format, occupy 3.8 TB.

In order to avoid storing all the decompressed images, one would have to perform an \enquote{online} decompression asynchronously coupled with an \enquote{online} mini-batch build, in order to feed the neural network \enquote{on flight}. Note that with such a solution it could be possible to accelerate the global learning time, by directly working with the GPU RAM, instead of the CPU RAM or the Hard Disk, which have longer access time. This \enquote{online} treatment is not an easy task to carry, 
and the problem will have to be addressed for databases exceeding tens or even hundreds of million images.

The training set is split into two sets with 90\% for the \enquote{real} training set and 10\% for validation. As said previously, the test set 
is always the same and is made of 200k images (cover and stegos).
		\subsection{Hyper-parameters}
To train our CNN we used a mini-batch stochastic gradient descent without dropout. We used the majority of the hyper-parameters of the article \cite{Huang2019_LMC}. The learning rate, for all parameters, was set to 0.002 and is decreased at the epoch 130 and 230, with a factor equal to 0.1. The optimizer is Adam, and the weight decay is 5e-4. The batch size is set to 100 which corresponds to 50 cover/stego pairs. In order to improve the CNN generalization, we shuffled the entire training database at the beginning of each epoch. First, layer was initialized with the 30 basic high-pass SRM filters, without normalization, and the threshold of the TLU layer equals 31 similarly to \cite{Ye2017}, \cite{Yedroudj2018_Net}. We made an early stop after 250 epochs as in \cite{Huang2019_LMC}. The code and all materials are available at the following link: http://www.lirmm.fr/\~{}chaumont/LSSD.html
		\subsection{Results and discussion}
		\label{sec:exp:results}
The different learning sets, from 20k to 1M images (covers and stegos), were used to test the LC-Net. Table \ref{tab:results} gives the network performances when tested on the 200k test cover/stego images database. Note that several tests were conducted for each size of the learning set and the displayed accuracies represent an average computed on the 5 best models recorded during the training phase. Those 5 best models are selected thanks to the validation set.

\begin{table}[!h]
	\centering
	
	\caption{Average accuracy evaluated on the 200k cover/stego images test set, with respect to the size of the learning database. }
	\label{tab:results}
	
	\begin{tabular}{|c|c|c|c|c|}
		\hline
		\textbf{Database size} & \textbf{Nb. of tests} & \textbf{Accuracy} & \textbf{Std. dev.} & \textbf{duration}\\
		\hline
		\hline
		20,000 & 5 & 62.33\% & 0.84\% & 2h 21 \\
		\hline
		100,000 & 5 & 64.78\% & 0.54\% & 11h 45 \\
		\hline
		200,000 & 5 & 65.99\% & 0.09\% & 23h 53 \\
		\hline
		1,000,000 & 1 & 68.31\% & / & 10d to 22d 
		 \\
		\hline
	\end{tabular}

\end{table}

\begin{figure}[h!]
	\centering

	\includegraphics[scale=0.7]{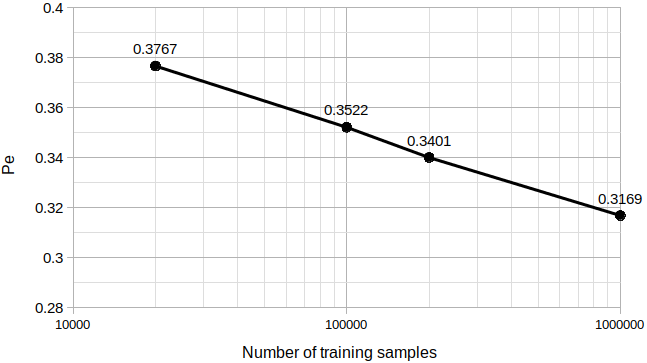}
	\caption{Average probability of error with respect to the learning database size. Notice that the abscissa scale is logarithmic.}
	\label{fig:results:curve_acc}
\end{figure}

Before analyzing the network results, we should note that learning times become significant (from 10 days to 22 days~\footnote{Experiments with 1M images were disrupted by a maintenance of the platform and it took 22 days. Nevertheless, we rerun the experiment on a similar platform, without any disruption in learning, and the duration was only 10 days. So we report both values to be more precise.}) once the number of images exceeds 1 million. This is a severe problem which did not allow us, due to lack of time, to run an evaluation on the 2 million (1M covers + 1M stegos) and 4 million (2M covers + 2M stegos) databases as the learning duration would have been higher than one month. Moreover, the hardware architecture would probably not be able to store all the database in RAM, and it is likely that the paging optimizations would be no longer valid on such a volume of data. The transfer of an image from hard disk to the RAM, and then to the GPU, becomes then the bottleneck in the learning process. As explained previously, to cope with this problem, a decompression thread and a reading thread with the use of a shared file and the use of semaphores could be used to reduce the memory storage and the transfer on the GPU memory. It would make it possible to build image mini-batches \enquote{on flight} during network learning.

Results of Table \ref{tab:results}, obtained for the payload 0.2 bpnzacs, confirm that the larger the learning set is (100k, 200k, 1M), the better the accuracy is. For the 20k database, the accuracy is $62\%$ and increases by almost $2\%$ each time the size of the learning set is increased. Moreover, the standard deviation is getting smaller and smaller, which highlights that the learning process is more and more stable as the database increases.

These first results would mean that most of the steganalysis experiments run by the community, using a medium size (but also a large size) Deep Learning network, are not done with enough examples to reach the optimal performance, since most of the time the database is around 10k (BOSS learning set) to 150k images (Alaska\#2 learning set with only one embedding algorithm). As an example, in our experiments, the accuracy is already improved by 6\% when the database increase from 20k to 1M images and the accuracy can probably be improved by increasing the dataset size since the irreducible error region is probably not reached.

Those results also confirm that a medium-size network such as LC-Net does not have its performance collapsing when the database size increases.

More interestingly, we observe (see Figure \ref{fig:results:curve_acc}) an exponential decrease in the probability of error in the function of the dataset size. This is a direct observation of the power-law discussed in Section \ref{sec:model_data_scaling}. Using a non-linear regression with Lagrange multipliers\footnote{The initial point for the non-linear regression is set to $a'=0.5$, $\alpha'=0.001$ and $c'_\infty = 0.01$, with $c'_\infty$ forced to be positive. The Matlab function is {\it fconmin} and the stop criterion is such that the mean of the sum of square error is under $10^{-6}$.}, we have estimated the parameters of Equation \ref{eq::power-law-dataset-size}: 

\begin{equation}
\epsilon(n) = 0.492415 n^{-0.086236}+0.168059
\label{eq::power-law-dataset-size-fit}
\end{equation}

The sum of the square error is $4.4\times10^{-6}$. Since there are only four points for the regression, it is probably erroneous to affirm that the irreducible error is $c'_\infty=16.8\%$. However, we can use equation \ref{eq::power-law-dataset-size-fit} to predict without much error, that if we use 2M images for the learning, the probability of error will be close to 30.9\%, and if we use 20M images, the probability of error will be again reduced of 2\% and will be 28.3\%. Note that if $c_\infty$ was equal to 0, the probability of error would be 30.7\% for 2M images, and 27.8\% for 20M images. If we consider a probability of error of $28.3$\% for 20M of images, the gain obtained compared to the probability of error  of $37.7\%$ with 20k images, corresponds to 9\% increase which is a considerable improvement in steganalysis domain.


To conclude, the error power-law also stands for steganalysis with Deep Learning, and this even when the networks are not very big (300,000 parameters), even when starting with a medium-size database (here, only 20k images), and even if the database is diverse (use of Alaska\#2 development script and around 100 camera models). So, bigger databases are needed for optimal learning, and using more than one million images are likely needed before reaching the \enquote{irreducible error} region \cite{Hestness2017}.


	\section{Conclusion}


In this paper, we first have recalled the recent results obtained by the community working on AI, and related to the behaviour of Deep-Learning networks when the model size or the database size is increasing. We then proposed an experimental setup in order to evaluate the behaviour of a medium-size CNN steganalyzer (LC-Net) when the database size is scaled. 

The obtained results show that a medium-size network does not collapse when the database size is increased, even if the database is diverse. Moreover, its performances are increased with the database size scaling. Finally, we observed that the error power-law is also valid for steganalysis domain. We thus estimated what would be the accuracy of the network if the database would have been made of 20 million images.

Future work will require to be done on a more diverse database (quality factors, payload size, embedding algorithm, colour, less controlled database), and also other networks. More practically, an effort should be made in order to reduce the learning time, and especially memory management. Finally, there are still open questions to solve such as: finding a more precise irreducible error value, finding the slope of the power-law depending on the starting point of the CNN (use of transfer, use of curriculum, use of data-augmentation such as pixels-off \cite{Yedroudj_pixelsoff}), or finding innovative techniques when the database is not huge in order to increase the performances.

	\section*{Acknowledgment}
The authors would like to thank the French Defense Procurement Agency (DGA) for its support through the ANR Alaska project (ANR-18-ASTR-0009). We also thank IBM Montpellier and the Institute for Development and Resources in Intensive Scientific Computing (IDRISS/CNRS) for providing us access to High-Performance Computing resources.


%
%
\bibliographystyle{splncs04}
\bibliography{bibliography}
\end{document}